\documentclass[%
 aip,
 amsmath,amssymb,
 reprint,%
]{revtex4-1}

\pdfoutput=1
\usepackage{graphicx}
\usepackage{dcolumn}
\usepackage{bm}

\usepackage{xcolor}
\usepackage[normalem]{ulem}

\usepackage[utf8]{inputenc}
\usepackage[T1]{fontenc}
\usepackage{mathptmx}
\usepackage{etoolbox}
\usepackage{siunitx}

\begin{document}


\title[]{Two-dimensional Thomson scattering in high-repetition-rate laser-plasma experiments}

\author{H. Zhang}
\affiliation{Department of Physics and Astronomy, University of California - Los Angeles, Los Angeles, California 90095, USA}

\author{J.J. Pilgram}
\affiliation{Department of Physics and Astronomy, University of California - Los Angeles, Los Angeles, California 90095, USA}

\author{C.G. Constantin}
\affiliation{Department of Physics and Astronomy, University of California - Los Angeles, Los Angeles, California 90095, USA}

\author{L. Rovige}
\affiliation{Department of Physics and Astronomy, University of California - Los Angeles, Los Angeles, California 90095, USA}

\author{P.V. Heuer}
\affiliation{Laboratory for Laser Energetics, University of Rochester, Rochester, NY, 14623, USA}

\author{S. Ghazaryan}
\affiliation{Department of Physics and Astronomy, University of California - Los Angeles, Los Angeles, California 90095, USA}

\author{M. Kaloyan}
\affiliation{Department of Physics and Astronomy, University of California - Los Angeles, Los Angeles, California 90095, USA}

\author{R.S. Dorst}
\affiliation{Department of Physics and Astronomy, University of California - Los Angeles, Los Angeles, California 90095, USA}

\author{D.B. Schaeffer}
\affiliation{Department of Physics and Astronomy, University of California - Los Angeles, Los Angeles, California 90095, USA}

\author{C. Niemann}
\affiliation{Department of Physics and Astronomy, University of California - Los Angeles, Los Angeles, California 90095, USA}

\date{\today}

\begin{abstract}
We present the first two-dimensional (2D) optical Thomson scattering measurements of electron density and temperature in laser-produced plasmas. The novel instrument directly measures $n_e(x,y)$ and $T_e(x,y)$ in two dimensions over large spatial regions (cm$^2$) with sub-mm spatial resolution, by automatically translating the scattering volume while the plasma is produced repeatedly by irradiating a solid target with a high-repetition-rate laser beam (10 J, $\sim$10$^{12}$ W/cm$^2$, 1~Hz). In this paper, we describe the design and auto-alignment of the instrument, and the computerized fitting algorithm of the spectral distribution function to large data-sets of measured scattering spectra, as they transition from the collective to the non-collective regime with distance from the target. As an example, we present 2D scattering measurements in laser driven shock waves in ambient nitrogen gas at a pressure of 95~mTorr. 

\end{abstract}

\maketitle

\section{\label{sec:level1}Introduction}

Thomson Scattering (TS) is a powerful, first-principles technique that is extensively used in plasma diagnostics to obtain electron temperatures and densities \cite{evans}. As the scattered spectra are collected along the path of the incident probe beam, one-dimensional (1D) Thomson scattering is done on a regular basis by imaging along the Rayleigh length of the probe beam \cite{gregori2004}. In order to probe across a  multi-dimensional volume or plane, common strategies include using multiple laser cords \cite{casey1992}, single laser beam probing over small regions (well within beam width), and repeat measurements at multiple locations \cite{tomita2012}.

For experiments in which the plasma profile is reproducible, the process of probing different regions of the plasma can be exploited by automatically controlling motions of the laser beam path and collection optics to continuously take series of measurements over a plane or volume, which not only enables implementation of two-dimensional (2D) Thomson scattering measurements with a simple one-laser setup, but also significantly reduces measurement time. This increase in efficiency is crucial for continuous measurements over very large sets of configurations, because prolonged experiments and repeated manual change of setup are often sources of larger errors.

Measurements of the spatial variation of density and temperature in laser-produced plasmas (LPP) are crucial for the interpretation of laboratory experiments on perpendicular \cite{niemann2014, schaeffer2014} and parallel \cite{weidl2016, heuer_par} collisionless shocks \cite{schaeffer2017,schaeffer2019} and related instabilities \cite{heuer2018}, diamagnetic cavities \cite{winske2019}, magnetic reconnection \cite{gekelman2008}, collisionless momentum transfer \cite{bondarenko2017, dorst2022}, artificial magnetospheres \cite{schaeffer2022}, or the generation of spontaneous magnetic fields via the Biermann battery \cite{pilgram}. However, with conventional Thomson scattering setups, only limited data points can be obtained for each configuration of the experiment. Here, we present a novel technique with programmed motions of optical components and probes mounted on stepper-motor controlled stages that implements two-dimensional Thomson scattering in high-repetition rate LPP. The technique is enabled by an optical design that maximizes alignment stability and optical extent \cite{kaloyan2022}, a triple grating spectrometer with notch filter for stray-light suppression \cite{ghazaryan2022}, a motorized auto-align scattering system \cite{kaloyan2021}, as well as a laser-target and data-acquisition system that produces LPPs at high repetition rate \cite{schaeffer2018}. This new technique significantly increases efficiency and spatial resolution of Thomson scattering spectra collection over a full planar region of the LPP. Combined with automated data processing and spectra fitting with the python {\it PlasmaPy} package \cite{plasmapy}, our approach effectively enables for the first time a high-resolution, single-beam, two-dimensional Thomson scattering measurement in laser-plasmas.
This capability to collect volumetric measurements of plasma density and temperature is of interest to a new generation of high-repetition-rate high-power laser facilities \cite{eli, hrr1, haefner, hrr2, hrr3}, where it could help accelerate the advancement of high-energy density science \cite{Ma_2021}. 

\section{Experimental Setup}
\begin{figure}
    \centering
    \includegraphics[width=1\linewidth]{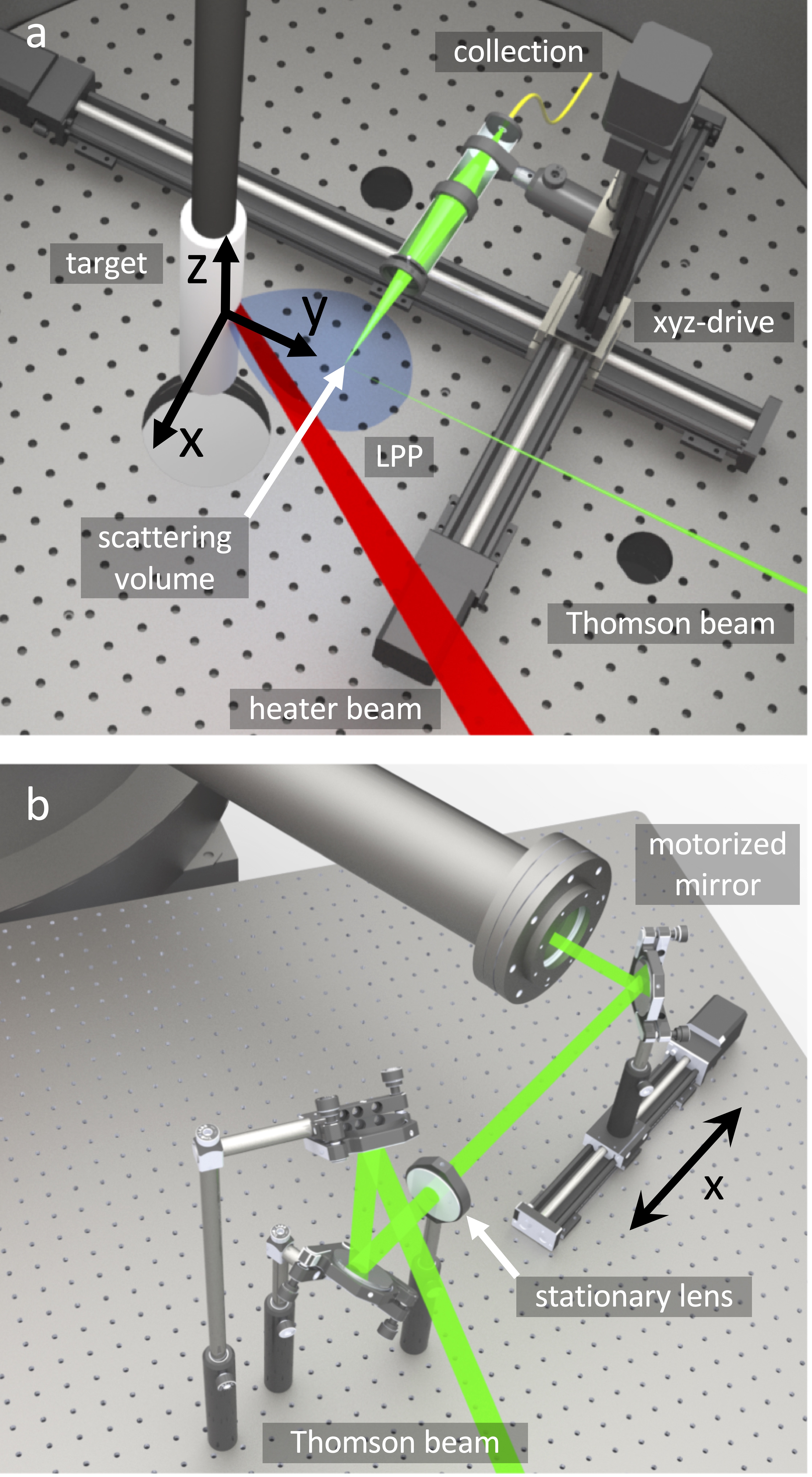}
    \caption{
    {(a) Schematic layout of the experimental setup showing the target and beam configuration. The LPP explodes along the $\hat{y}$-axis. The collection lens and scattering volume are automatically aligned using a motorized xyz-stage inside the vacuum chamber. (b) The Thomson probe beam can be translated along the $\hat{x}$-axis using a motorized turning mirror. Translating the probe beam and collection lens simultaneously allows 2D measurements in the xy-plane.}
    }
    \label{fig:setup}
\end{figure}

Experiments were performed at the {\it Phoenix} laser facility at the University of California Los Angeles \cite{niemann2012} using a flashlamp pumped high-average power laser with Nd:glass regenerative zig-zag slab-amplifier and  wavefront correction by stimulated Brillouin scattering \cite{dane}.
A schematic of the experimental setup is shown in figure \ref{fig:setup}. 
The plasma is generated at 1~$\si{\hertz}$ repetition rate by ablating a 38~mm diameter cylindrical high-density polyethylene (C$_2$H$_4$) target with a 10~$\si{\joule}$, 1053~$\si{\nano\meter}$ laser pulse at 20~$\si{\nano\second}$ full-width at half maximum (FWHM) duration. 
The beam energy is stable to within 5$\%$, and the pulse shape and pointing are stable to within 1$\%$.
An f/26 spherical lens outside the vacuum chamber is used to focus the heater beam  
to a 250~$\si{\micro\meter}$ focal spot on the target, resulting in an intensity around 10$^{12}$ W/cm$^2$.
While the beam is incident at an angle of 34$\si{\degree}$ relative to the normal, the plasma explodes perpendicularly to the target surface and along the $\hat{y}$-axis. To ensure that the beam ablates a fresh region on the target in each shot, the target is controlled to rotate and translate in a helical pattern at 1~Hz, accommodating around 7,000 total shots per target. 

A separate  beam ($\lambda_i=532$~$\si{\nano\meter}$) from a frequency doubled Nd:{YAG} laser with 4~$\si{\nano\second}$ (FWHM) duration and 460~$\si{\milli\joule}$ pulse energy at 20~$\si{\hertz}$ repetition-rate is used as the Thomson scattering probe. The probe beam is injected parallel to the plasma blow-off axis ($\hat{y}$-axis), is linearly polarized along the $\hat{z}$-axis using a waveplate,
and is synchronized to the heater pulse, including a variable delay.
A stationary spherical lens outside the vacuum chamber with a focal length of $f=1.5$~m is used to focus the probe beam to a 200~$\si{\micro\meter}$ spot at a distance of 2~cm from the target surface. The slow f/150 focusing lens results in a large $\pm$3~cm Rayleigh range, creating a cylindrical beam focus of constant diameter over the entire region of interest. 
%
The probe beam terminates on the target, creating a secondary laser-plasma as well as 
copious amounts of stray light. 
The secondary plasma plume reaches the scattering volume well after the time of the scattering measurement and consequently does not affect the results.

Scattered light is collected perpendicular to the probe beam (scattering angle $\theta=90^{\circ}$) into an f/20 collection cone by a 5.0~cm focal length, 12~mm diameter aspheric collection lens and is focused at f/5.3 into a 200~$\si{\micro\meter}$ core optical fiber with a numerical aperture of $NA=0.22$. 
The collection lens projects the fiber with a 3.8x magnification onto the probe beam so that the 0.76~mm diameter fiber core projection exceeds the probe beam width by a factor of 3-4. 
This optical design with a relatively small collection angle is used to minimize alignment sensitivity, accommodate a range of probe beam diameters, and to assure that alignment between the beam and fiber projection is maintained during a motorized raster scan over a large area, and for the duration of an experimental run lasting several hours. 
%
The scattered light collection efficiency is best evaluated using the optical extent $G =  A'\Omega' = A\Omega$, which is the same for the source and detector \cite{kaloyan2022}. Here, $\Omega'$ is the collection solid angle and $A'$ is the source area defined by the intersection of the circular fiber projection and the pencil beam. Similarly, $A$ and $\Omega$ are the fiber core area and solid angle of light entering the fiber and detector, respectively.  
The parameters used here result in $G'$= 3.0$\times$10$^{-4}$ sr$\cdot$ mm$^2$, which is 
35$\%$ of the 
theoretical upper limit for the throughput $G_{max}$ = 8.8$\times$10$^{-4}$~sr$\cdot$mm$^2$, determined by the spectrometer acceptance cone and fiber area.
The collection lens, disposable blast shield, 5~mm clear-aperture iris diaphragm, and fiber launch are assembled into a 25~mm diameter light-tight metal tube, which is mounted onto an xyz-motorized stage inside the vacuum chamber. The stepper-motors can be controlled with 5~$\si{\micro\meter}$ resolution and bidirectional repeatability.

A stepper-motor-controlled turning mirror outside the laser entrance port translates the probe beam transverse to the blow-off axis ($\hat{x})$ in order to shift the scattering volume horizontally (Fig. \ref{fig:setup}b). 
The scattering volume is defined by the intersection of the probe beam and the fiber-projection, resulting in a spatial resolution of 0.2$\times$0.7~mm$^2$. Translating the beam and collection branch simultaneously allows the scattering volume to be moved anywhere within the horizontal plane.
Keeping the lens stationary assures that the beam remains exactly parallel to the blow-off axis when translating the turning mirror. 
Shifting the beam along $\hat{x}$ then offsets the focus along $\hat{y}$ by an equal distance. However, in this setup the range is limited to $\Delta x \in$ \{-2~cm, +2~cm\}, which is well within the depth of focus.

Scattered light collected by the fiber is coupled via a vacuum feedthrough to the input slit of a triple-grating f/4 Czerny-Turner imaging spectrometer \cite{ghazaryan2021}. 
The first two spectrometer stages are configured as a double-subtractive system to act as a tuneable notch filter that blocks the unshifted probe laser stray light around 532~nm. 
The third stage disperses the stray light subtracted spectrum onto the detector.
Holographic gratings, in combination with baffles, and an intermediate slit between the second and third stages are used to minimize the stray light.
The three 110$\times$110$~\si{\milli\meter^2}$ aluminum-coated, 1200~grooves/mm gratings are blazed at 500~nm for an efficiency around 60$\%$ at 532~nm. 
The notch filter blocks stray light over a 1.5~nm range by four orders of magnitude \cite{ghazaryan2022}.

Spectra are captured by an image intensified charged coupled device (ICCD) equipped with a generation III photocathode with a quantum efficiency of around 50\% at 532~nm. The microchannel plate is gated at 4~ns at the maximum gain. Despite using 2 $\times$ 2 hardware binning the maximum pixel count is only a small fraction of the 16-bits maximum and well within the linear response range. For all data presented here, the spectra were further software binned over all 512 vertical pixels into 512 total bins with 0.0388~nm/bin. To maximize throughput, both the input and intermediate slits were fully open (4~mm), and so the fiber core itself serves as a 0.20~mm wide slit. The measured instrument function (not shown) is Gaussian with a width of 0.29 nm (FWHM) over a spectral range of 19.8 nm.

Plasma self emission lines are also present in the spectrum and can have intensities comparable to the TS signal. 
Two fast solenoid shutters are therefore used for interleaving an equal number of TS images and plasma-only images for background subtraction. 
The first shutter is synchronized to the probe laser Q-switch and reduces the probe beam frequency to 1~Hz. The second shutter is synchronized to the data-acquisition system and pulses the probe beam on for only every other laser-target shot so that a background image is recorded for every TS spectrum back-to-back.

\section{Results and discussion}
\begin{figure}
    \centering
    \includegraphics[width=1\linewidth]{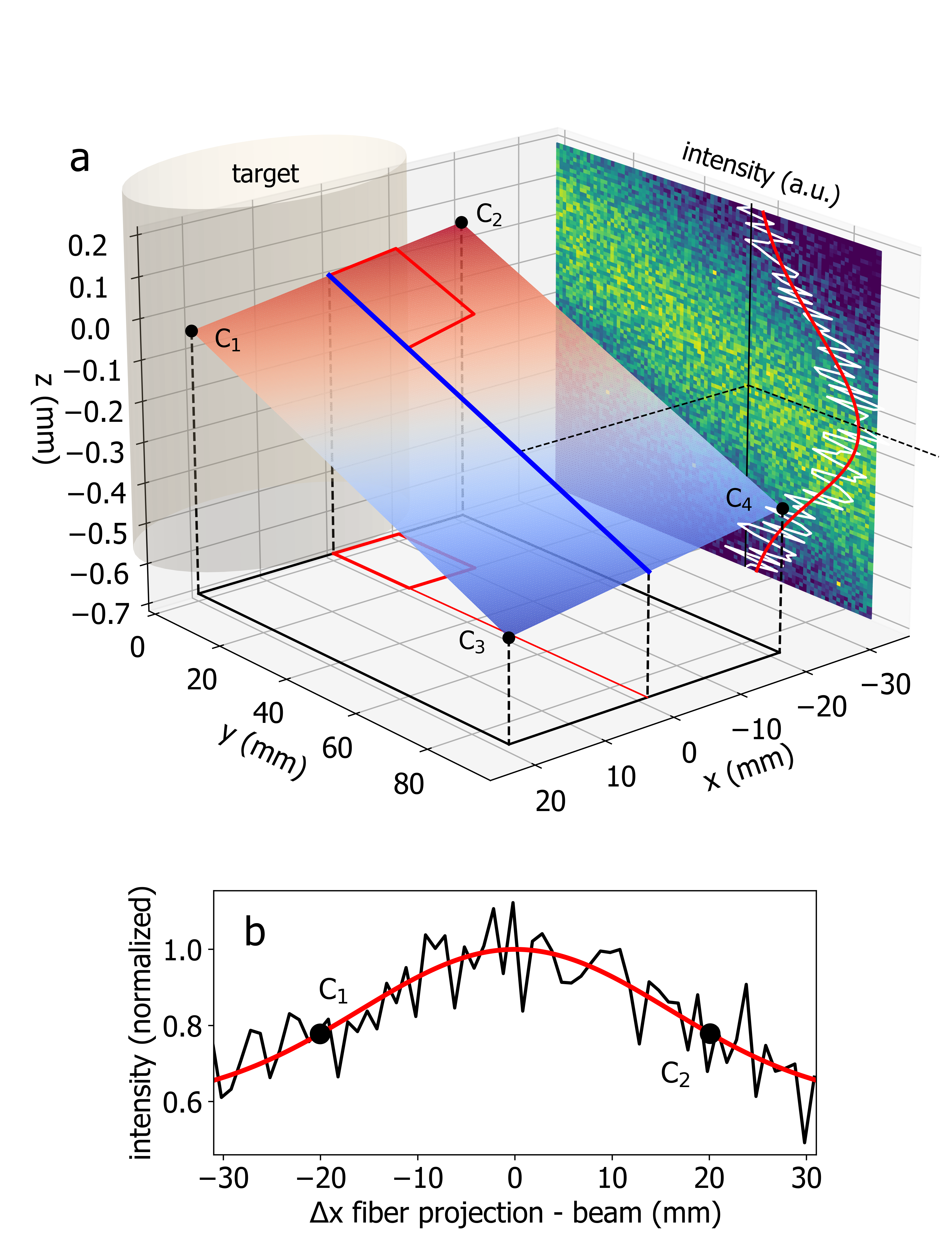}
    \caption{
    (a) Variation of the probe beam height $z(x,y)$ as interpolated from Rayleigh scattering measurements of the beam location at the four corners $C_1$-$C_4$.  The plane is inclined since the probe beam points up slightly at an angle of 0.4$^{\circ}$. The red box indicates the much smaller plane used for the TS measurements in figures \ref{fig:selfemission} and \ref{fig:2dts}. The green contour of a $90\times90$ pixel vertical raster scan of Rayleigh scattering also shows the tilted probe beam at $x=0$ (blue line). The z-lineout at $y=55$~mm (white curve) is consistent with a Gaussian with a width of $\sigma=0.27$~mm. (b) Variation of the collection efficiency with longitudinal fiber alignment, measured by translating only the collection assembly along $\hat{x}$ while keeping the probe laser fixed at $x=0$ (blue line in Fig. a). }
    \label{fig:autoalign}
\end{figure}

Before conducting an automated raster scan of the Thomson scattering signal, the probe beam height $z(x,y)$ and inclination in the region of interest are mapped out using Rayleigh scattering off of nitrogen gas at pressures around 100~Torr. 
Figure \ref{fig:autoalign} shows the probe beam plane as measured in the experiment over the full range of travel $y \in$ \{0, 90~mm\} and $x \in$ \{-20~mm, +20~mm\}. 
The plane is inclined since the probe beam points up slightly at an angle of 0.4$^{\circ}$. The plane is not shown to scale but is stretched vertically by a factor of 100. 
The green filled contour projected onto the right pane of Fig. \ref{fig:autoalign}a is a 90$\times$90 raster scan of the total measured Rayleigh scattering intensity $I(y,z)$. The data was obtained by keeping the probe beam fixed at $x=0$ (blue line), while scanning the collection assembly in $y$ and $z$ in 1~mm $\times$ 0.01~mm steps.
The $z$-lineout of that data at $y=55$~mm (white curve) is consistent with a Gaussian (red fit) with a width of $\sigma=0.28$~mm. The width remains constant over the entire range within the accuracy of the measurement, consistent with the large Rayleigh range of the beam focus.
%
%
In order to construct a plane and motion-profile for a raster Thomson scan, the beam height $z(x,y)$ is interpolated from Rayleigh scattering measurements at the four corners C$_1$ through C$_4$. For these measurements, the beam is first translated to the required $x$ using the final turning mirror, while the collection assembly is  moved to the required position $(x,y)$. The Rayleigh scattering intensity $I(z)$ is then measured while scanning across the beam at 1~Hz in 0.1~mm steps, and the beam center $z$ is determined from a Gaussian fit to the experimental data. Measuring the $z$-coordinates of the four corners $z_1$ through $z_4$ allows us to calculate the beam height at an arbitrary position in the plane. If the four corners have the coordinates $C_1=(x_0,0),\ C_2=(-x_0,0),\ C_3=(x_0,y_0),\ C_4=(-x_0,y_0)$, then the $z$-coordinate of the beam position at $(x,y)$ is given by the following formula:
\begin{equation}
    \begin{split}
        z(x,y)=(1-\frac{x_0-x}{2x_0})(1-\frac{y}{y_0})z_1+\frac{x_0-x}{2x_0}(1-\frac{y}{y_0})z_2 \\
        +\frac{y}{y_0}(1-\frac{x_0-x}{2x_0})z_3+\frac{(x_0-x)y}{2x_0y_0}z_4\ .
    \end{split}
\end{equation}
Inclusion of all four corner coordinates in the formula accounts for possible change of beam inclination angle over the range of $x$ coordinates. The projection of such a "twisted" plane onto the $xy$-plane is an equidistant grid. 
Figure \ref{fig:autoalign}b shows the measured Rayleigh scattering intensity as a function of the offset $\Delta x$ between the fiber projection and the probe-beam. For this measurement the beam is kept stationary at $x=0$ while the collection assembly is scanned along $x$ at a constant $y=0$, and $z=0$. The figure shows that keeping the collection lens stationary while only moving the beam will result in a 20$\%$ drop in collection efficiency from the center of the plane to each corner. In the raster scans the beam and collection lens are therefore always moved in synchronization ($\Delta x=0$). However, 2D-scans of smaller planes of only a few millimeters in size, could be accomplished without motion of the collection lens along $\hat{x}$.
This technique is also used during the initial alignment to find the optimum stand-off between the collection lens and the probe beam (i.e. the peak of the curve in Fig. \ref{fig:autoalign}b).

To demonstrate this technique, 
two-dimensional Thomson scattering was used to map out the density and temperature in a shock wave driven by a laser-plasma exploding into ambient gas.
Shocks produced by explosive outflows (blast waves) have been studied in numerous laser-plasma experiments \cite{Grun1991Instability, Edens2010Study} to help understand shocks in space and astrophysical systems \cite{McKee1991Interstellar, vernstrom}. 
Two-dimensional TS measurements in blast waves are of particular interest, due to the important role of misaligned density and temperature gradients in such shock-waves in generating cosmic magnetic fields via the Biermann battery effect \cite{Gregori2012Generation}. 
In the experiments, the target chamber is pumped to $2\times10^{-5}~\si{Torr}$ using a turbomolecular pump before back-filling with nitrogen gas. A dynamic equilibrium of ($95 \pm 3$)~mTorr is maintained during 1~Hz laser-target ablation as measured by a capacitive manometer.
Thomson scattering measurements
were performed at $t=(100 \pm 5)$~ns after the heater pulse, measured peak-to-peak, over a planar region with range of coordinates of $x \in \{-10,0\}~\si{\milli\meter}$  and  $y\in \{4, 23\} ~\si{\milli\meter}$, vertically aligned with the probe beam. 
The resolution was 0.5$~\si{\milli\meter}$, which corresponds to 20$\times$38 data points. The reason that x-coordinates are chosen to be negative is purely practical: shock waves are symmetric about the blow-off axis; aligning the Thomson diagnostic laser to positive $x$-coordinates would cause it to irradiate an area of the target that is to be ablated by the heater beam in subsequent shots, which will then interfere with plasma generation. Spectra at each coordinate are averaged over five individual shots and subtracted by five plasma-only background shots without the probe beam.

\begin{figure}
    \centering
    \includegraphics[width=1.\linewidth]{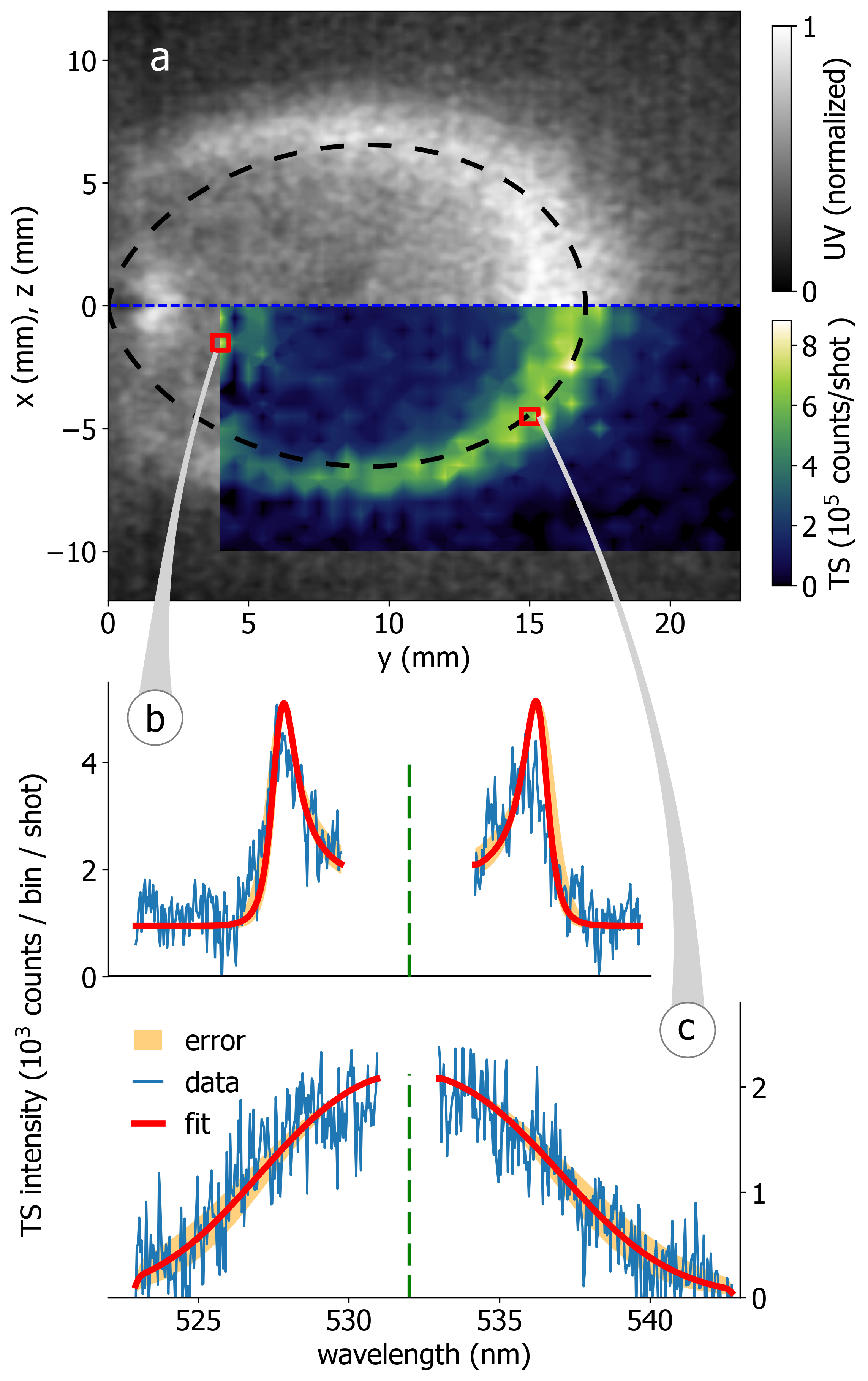}
    \caption{(a) The composite image of a laser-driven blast wave at t=100~ns after the heater pulse combines the UV self-emission recorded with a gated imaging camera, and the TS signal intensity constructed from a 38$\times$21 raster scan. The black segmented line corresponds to $r\sim \cos^2(\Theta)$. Self emission images show that the shock wave is symmetrical about the blow-off axis. (b) A typical collective TS spectrum near the target, averaged over five shots and five plasma-only background shots. The fit (red curve) is consistent with $T_e=2.3\pm0.3$~eV and $n_e=1.3\times10^{17}$~cm$^{-3}$, corresponding to $\alpha=2.0$. (c) A typical spectrum inside the blast wave that is essentially non-collective at $T_e=14.5\pm 4.0$~eV and $n_e=2.5\times10^{16}$~cm$^{-3}$, corresponding to $\alpha=0.34$\ . The shaded orange regions indicate the sensitivity of the fits to temperature and correspond to 20$\%$ variation in $\chi^2$.
    }
    \label{fig:selfemission}
\end{figure}

A second ICCD with a generation II intensifier sensitive in the ultraviolet (UV) range with a 4~ns gate width and a 25~mm focal length objective was used to image the plasma and blast wave in the $yz$-plane from the side.
Figure \ref{fig:selfemission}a overlays the integrated Thomson scattering signal intensity measured in a horizontal $(x,y)$ plane and the plasma self-emission images
at $t=(100 \pm 5)$~ns after the heater pulse.
The blast wave is clearly visible in both the TS and the self-emission images at a radius around 16~mm.
The elliptical shape is consistent with a typical initial angular velocity distribution of laser-produced plasmas of $v\sim \cos^2(\Theta)$, as indicated by the segmented black line \cite{heuer2017}, where $\Theta$ is the angle relative to the blow-off axis.
The cos$^2$-scaling overestimates the shock wave position for small $\Theta$ because the shock slows with distance from the target. 
 We note that the self-emission images and TS data represent two orthogonal planes. However, the blast wave is cylindrically symmetric about the blow-off axis.
Thomson scattering can be collective or non-collective \cite{froula12}, depending on the dimensionless scattering parameter $\alpha = 1 /(k \lambda_D)$, where $k= 4\pi \cdot sin(\theta/2)/\lambda_i$ is the scattering vector and $\lambda_D = \sqrt{\epsilon_0 k_B T_e/n_e e^2}$ is the Debye length.
When $\alpha > 1$, the scattering is collective, and the spectrum is dominated by scattering off of plasma waves. The spectrum is then non-Gaussian, even for a Maxwellian velocity distribution function as assumed here. For $\alpha \ll 1$, the scale length of electron density fluctuation sampled by TS is small compared to the Debye length, and
the scattering is non-collective. The spectrum then has a Gaussian shape if the underlying velocity distribution function is Maxwellian. 
%
The spectral density is given by the equation
\begin{equation}
    S(k,\omega) = \sum_e \frac{2\pi}{k} \left| 1 - \frac{\chi_e}{\epsilon}\right|^2 f_e \left( \frac{\omega}{k}\right) +
    \sum_i \frac{2\pi Z_i}{k} f_i\left(\frac{\omega}{k}\right)\ ,
\end{equation}
where $Z_i$ is the ion charge state, $\omega$ is the scattering frequency, $f_e$ and $f_i$ are the particle velocity distributions for the electrons and ions,  $\epsilon = 1 + \sum_e \chi_e + \sum_i \chi_i$ is the dielectic function, and $\chi_e$ and $\chi_i$ are the electron and ion components of the susceptibility, respectively. 

Near the target, where the densities are high and the temperatures are low, the scattering is collective. A spectrum measured at $(y,x)$ = (4~mm, -1.5~mm) is shown in Figure \ref{fig:selfemission}b. Only data that is not blocked by the notch around $\lambda_i$ (green segmented line) and that is used for the fit is shown in the graph.
In this regime, the spectrum is dominated by two wings on either side of $\lambda_i$ caused by scattering off forward and reverse-propagating electron plasma waves \cite{schaeffer2016}. The separation of the wings depends on $n_e$ and their width on $T_e$. Both the density and temperature can therefore be determined explicitly from a single spectrum.
The data agrees well with the spectral density function for $T_e=2.1$~eV and $n_e=1.3\times10^{17}$~cm$^{-3}$, corresponding to $\alpha=2.0$, and consistent with simulations for similar laser-plasma parameters \cite{schaeffer_lpp}.
Stray light is effectively filtered out by the notch, except for measurements this close to the target ($y<6$~mm), where the target surface is within the direct view of the fiber. Residual stray light leaking through the wings of the notch can then be as much as an order of magnitude larger in amplitude than the TS signal. 
We use a slightly wider notch region of 4.4~nm in the fit to account for this leakage.
The offset in Fig. \ref{fig:selfemission}b observed very close to the target is also in part due to stray-light leaking through the notch and distributed over the entire spectrum \cite{ghazaryan2022}.


Farther from the target and inside the blast wave the spectrum is only weakly collective.
Figure \ref{fig:selfemission}c shows a TS spectrum measured at $(y,x)$ = (14~mm, -4.5~mm). The spectrum is essentially non-collective and nearly Gaussian in shape. In this regime, the temperature can be derived from the width of the broadened line and the density from the scattering signal intensity, after an absolute irradiance calibration \cite{ghazaryan2021}. Raman scattering off of nitrogen gas of a known density (at a pressure of 658~Torr) was used to absolutely calibrate the throughput of the optical system and spectrometer and determine the relation between the measured Thomson scattering signal $N_T$ and the density, $n_e = s\cdot N_T$, where $s=(3.11\pm 0.22)\times 10^{10}$~cm$^{-3}$/counts.  
At distances $y>6$~mm from the target the notch effectively filters all stray-light and only data within a 2~nm wider region around $\lambda_i$ is omitted from the fit.
The spectral density function that best fits the data in Fig. \ref{fig:selfemission}c is consistent with $T_e= 14.2$~eV and $n_e=2.5\times10^{16}$~cm$^{-3}$, corresponding to a scattering parameter of $\alpha=0.34$\ .
\begin{figure}[!hb]
    \centering
    \includegraphics[width=1\linewidth]{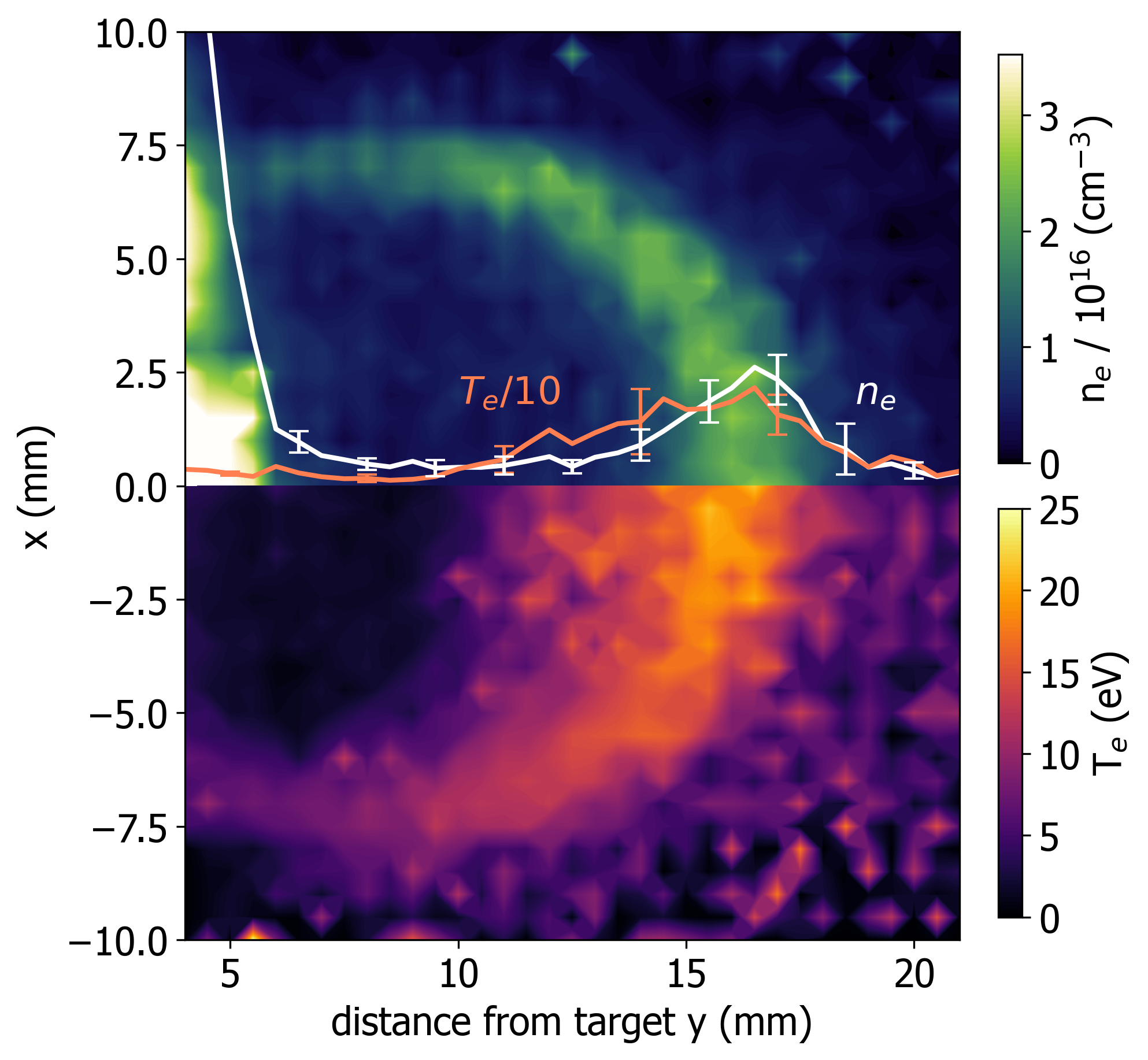}
    \caption{Two-dimensional Thomson scattering measurements of density and temperature in a laser-driven blast wave in 94~mTorr of nitrogen, t=100~ns after the heater pulse. The data was compiled by a 38$\times$21 raster scan over the course of five hours, averaging ten TS spectra per position (about 16,000 shots total). Data was only collected over half of the symmetric plane $x \in \{-10, 0\}$ mm. The density data is mirrored on the opposite half of the plane for a side-by-side comparison. The $T_e$ and $n_e$ profiles along $x=0$ show that the density falls of much faster behind the shock wave than the temperature.}
    \label{fig:2dts}
\end{figure}

In order to automatically process and fit the tens to hundreds of thousands of TS spectra produced in a typical high-repetition-rate experiment, the Thomson scattering fitting algorithm implemented in the python {
package \it PlasmaPy}~\cite{plasmapy} is used.
It includes a spectral density function that calculates Thomson scattering spectra assuming Maxwellian velocity distributions for given sets of plasma and instrument parameters including temperature, density, and instrument function. To perform a fit, the Thomson scattering parameters and range for target fit parameters are passed to a fit model function, which is then fitted using a nonlinear least squares fit algorithm \cite{lmfit} to determine $T_e$ and $n_e$ from the spectral density function.
For collective spectra, both temperature and density are obtained with the fit algorithm. An upper limit for density fit value of $10^{18}\si{\per\centi\meter\cubed}$ is used in order to avoid fitting to unrealistically high densities $>10^{19}\si{\per\centi\meter\cubed}$. For weakly collective spectra, density cannot be accurately determined from the spectra alone, so the algorithm first determines the density from the total signal intensity, as determined by the area under a Gaussian fit to the data, including also the spectrum blocked by the notch or the clipped wings outside the measured spectral range. Density is calculated from the total signal intensity via a Raman scattering calibration. The temperature is then determined from the fit to the spectral density function with $n_e$ as an input parameter. 
The {\it PlasmaPy} package provides a general, versatile approach to Thomson scattering data analysis, especially in fitting large data-sets, although manual verification of the fit results and tuning are still required for a few noisy spectra where the fit does not converge.

The first 2D TS measurements of density and temperature in a laser-driven blast wave in 94~mTorr of nitrogen, 100~ns after the heater beam, are shown in Fig. \ref{fig:2dts}.
The data was collected in a 38$\times$21 raster scan in 0.5~mm steps over the course of five hours, averaging ten TS spectra and ten background spectra per position (about 16,000 shots total). 
Density and temperature are determined as the average between two fits for each position to two sets of spectra, each obtained by averaging over signals collected at the same position in five consecutive shots. Representative errors for the fit temperature and density over the shots for the lineout at $x=0$ are also shown in the figure.
Data was collected only over half of the symmetric plane
$x \in$ \{-10~mm, 0\}. The density data is mirrored onto the opposite half of the plane for a side-by-side comparison. 
Measured temperatures and densities inside the shock wave are as high as 25~eV and $2.5\times10^{16}$~cm$^{-3}$.
The $T_e$ and $n_e$ profiles along $x=0$ show that the density falls off much faster behind the blast wave than the temperature. 

\section{Conclusion}
In this paper, we present the first 2D raster imaging TS measurements of density and temperature in a laser-produced plasma. The instrument measures the TS spectrum as a function of $x$ and $y$ over a horizontal plane by automatically translating the probe laser beam and collection optics between subsequent shots. 
Prior to TS data  collection, the height and inclination of the probe beam over the region of interest are mapped out via Rayleigh scattering off of nitrogen. This data provides input to then steer the collection lens in all three dimensions as needed.
The {\it PlasmaPy} TS spectral density model provides a robust algorithm for automatically retrieving the density and temperature from the tens of thousands of TS spectra collected in a typical high-repetition-rate run. 

In this experiment, the signal-to-noise ratio was limited by carbon and nitrogen emission lines with an amplitude comparable to the TS signal. TS spectra were therefore averaged over five shots (and five emission background shots) in order for the automated fit to converge.
A future upgrade to this diagnostic will increase the probe beam energy to 0.7~J. Scattering spectra with comparable SNR can then be produced in a single shot.
This technique can easily be expanded to three-dimensions (3D) for volumetric measurements by adding a second motorized turning mirror for the probe beam.
It could also be applied to map out the earlier stages of dense, millimeter-scale laser-plasmas  using piezo actuators with $\mu m$ repeatability, and a higher resolution spectrometer to resolve the ion feature \cite{glenzer1999}.

\section{Acknowledgements}
This work was supported by the National Nuclear Security Administration (NNSA) Center for Matter Under Extreme Conditions under Award Number DE-NA0003842, the Defense Threat Reduction agency (DTRA) and Livermore National Laboratory under contract number B655224, the Department of Energy (DOE) under award numbers DE-SC0019011 and DE-FC02-07ER54918, and the Naval Information Warfare Center-Pacific (NIWC) under contract NCRADA-NIWCPacific-19-354. This research made use of PlasmaPy version 2023.1.0, a community-developed open source Python package for plasma research and education (PlasmaPy Community et al. 2023). H.Z. was supported by the UCLA Department of Physics and Astronomy REU program. J.J.P was supported by the National Science Foundation Graduate Fellowship Research Program under award number DGE-1650604.  We thank Curtiss-Wright MIC for help with the slab laser system. We thank Zoltan Lucky and Tai Ly for their expert technical support.

\section*{Author declarations}

{\bf \noindent \small Conflict of Interest}\\
The authors have no conflicts to disclose.\\[10mm]

{\bf \noindent \small Author Contributions}\\

\noindent
{\bf H. Zhang:} Formal analysis (lead); Methodology (equal); Investigation (equal); Software (equal); Writing - original draft (equal); Writing - review $\&$ editing (equal). {\bf J.J. Pilgram: } Methodology (equal); Investigation (equal); Formal analysis (equal); Writing - review $\&$ editing (equal). {\bf C.G. Constantin:} Conceptualization (equal); Methodology (equal);  Investigation (equal); Writing - review $\&$ editing (equal). {\bf L. Rovige:} Investigation (equal); Writing - review $\&$ editing (equal). {\bf P.V. Heuer: } Formal analysis (equal); Software (equal); Writing - review $\&$ editing (equal). {\bf S. Ghazaryan:} Conceptualization (equal); Methodology (equal); Writing - review $\&$ editing (equal). {\bf M. Kaloyan:} Conceptualization (equal); Methodology (equal); Writing - review $\&$ editing (equal). {\bf R.S. Dorst:} Investigation (equal); Writing - review $\&$ editing (equal). {\bf D.B. Schaeffer: } Software (equal); Writing - review $\&$ editing (equal). {\bf C. Niemann: } Conceptualization (lead); Methodology (lead);  Investigation (equal); Formal analysis (equal); Writing - original draft (equal); Writing - review $\&$ editing (equal). \\

\section{DATA AVAILABILITY}
The data that support the findings of this study are available from the corresponding author
upon reasonable request.

\bibliographystyle{unsrtnat}
\bibliography{bibliography.bib}

\end{document}